\documentclass[pdftex,a4paper,12pt]{article}
\usepackage{amssymb,amsmath,amsfonts,ntheorem,nomencl,mathrsfs}
\usepackage[arrow, matrix, curve]{xy}
\usepackage[latin1]{inputenc}
\newcommand{\IC}{\mathbb{C}}
\newcommand{\IR}{\mathbb{R}}

\newcommand{\ILL}{\mathscr{L}}

\newcommand{\IHH}{\mathscr{H}}

\newcommand{\IPP}{\mathscr{P}}
\newcommand{\IFF}{\mathscr{F}}

\newcommand{\pa}{\slash\slash}

\newcommand{\Id}{{\rm d}}

\newcommand{\f}{\frac}
\newcommand{\nn}{\nonumber}

\newtheorem{theorem}{THEOREM}[section]

\newtheorem{Corollary}[theorem]{Corollary}

\newtheorem{Remark}[theorem]{Remark}

\newtheorem{Theorem}[theorem]{Theorem}

\newtheorem{Proposition}[theorem]{Proposition}

\newtheorem{Definition}[theorem]{Definition}

\begin{document}
\begin{titlepage}
\title{Kato's inequality and form boundedness of Kato potentials on arbitrary Riemannian manifolds}
  \author{Batu Güneysu\footnote{E-Mail: gueneysu@math.hu-berlin.de}\\
   Institut für Mathematik\\
   Humboldt-Universität zu Berlin\\
   }
\end{titlepage}

\maketitle 
\begin{abstract} 
Let $M$ be a Riemannian manifold and let $E\to M$ be a Hermitian vector bundle with a Hermitian covariant derivative $\nabla$. Furthermore, let $H(0)$ denote the Friedrichs extension of $\nabla^*\nabla/2$ and let $V:M\to\mathrm{End}(E)$ be a potential. We prove that if $V$ has a decomposition of the form $V=V_1-V_2$ with $V_j\geq 0$, $V_1$ locally integrable and $\max\sigma(V_2(\bullet))$ in the Kato class of $M$, then one can define the form sum $H(V):=H(0)\dotplus V$ in $\Gamma_{\mathrm{L}^2}(M,E)$ on arbitrary $M$. We also show that if $M$ is geodesically complete, then the smooth sections with compact support are a form core for $H(V)$.
\end{abstract}

\section{Introduction}

Let $M$ be a smooth Riemannian $m$-manifold\footnote{ We will only assume that $M$ is connected and without boundary.}, equipped with the Riemannian volume measure $\mathrm{vol}(\bullet)$ on the Borel sigma algebra $\mathscr{B}(M)$. The usual scalar Laplace Beltrami operator will be written as $\Delta=-\Id^*\Id$. Let $E\rightarrow  M$ be a smooth (finite dimensional) Hermitian vector bundle with a Hermitian covariant derivative $\nabla$. The symbol $\left\|\bullet\right\|_x$ stands for the norm and the operator norm corresponding to the Hermitian structure $(\bullet,\bullet)_x$ on each fiber $E_x$. For any section  $\Psi$ in $E$ or in $\mathrm{End}(E)$, we will use the notation 
\[
|\Psi|:M\longrightarrow [0,\infty),\>\>\left|\Psi\right|(x):=\left\|\Psi(x)\right\|_x. 
\]
The smooth sections in $E$ with compact support will be denoted with $\Gamma_{\mathrm{C}^{\infty}_0}(M,E)$, and $\Gamma_{\mathrm{L}^2}(M,E)$ stands for the Hilbert space of (equivalence classes of) measurable sections $f$ in $E$ such that
\[
\left\|f\right\|^2:=\int_M \left\|f(x)\right\|^2_x\mathrm{vol}(\Id x) <\infty,
\]
with scalar product 
\begin{align}
\left\langle f_1,f_2 \right\rangle =\int_M (f_1(x),f_2(x))_x \mathrm{vol}(\Id x).\label{xx3}
\end{align}                                                                                                                                                                           
To $\nabla$ and the Riemannian structure of $M$ there canonically corresponds the Bochner Laplacian
\begin{align}
\nabla^*\nabla:\Gamma_{\mathrm{C}^{\infty}_0}(M,E)\longrightarrow\Gamma_{\mathrm{C}^{\infty}_0}(M,E),\label{cci}
\end{align}
a second order elliptic differential operator that can be defined as follows: If $v_1,\dots,v_{m}$ is a local orthonormal frame near $x\in M$ for the tangential bundle $\mathrm{T} M$ (which will be considered as complexified) and if $\Psi\in \Gamma_{\mathrm{C}^{\infty}_0}(M,E)$, then 
\begin{align}
\nabla^*\nabla\Psi(x)=-\sum^{m}_{j=1} \nabla_{v_j}\nabla_{v_j} \Psi(x)+ \nabla_{\nabla^{\mathrm{T}M}_{v_j} v_j}\Psi(x).\label{s87}
\end{align}
Here, $\nabla^{\mathrm{T}M}$ stands for the Levi-Civita connection and (\ref{s87}) does not depend on the particular choice of the local orthonormal frame. Note that since $\nabla$ is compatible with $(\bullet,\bullet)_x$, the assignment (\ref{cci}) defines a symmetric nonnegative operator in $\Gamma_{\mathrm{L}^2}(M,E)$. Throughout, let $V:M\to\mathrm{End}(E)$ be a potential, that is, $V$ is a measurable section in $\mathrm{End}(E)$ such that $V(x):E_x\to E_x$ is self-adjoint for almost every (a.e.) $x\in M$. \vspace{1.2mm}

One of the most fundamental questions in nonrelativistic quantum physics is the proper definition of self-adjoint realizations in $\Gamma_{\mathrm{L}^2}(M,E)$ of operators that are formally given by $\nabla^*\nabla+V$. For example, operators that describe the energy of charged nonrelativistic spin $0$ particles can be written in the form $\nabla^*\nabla+V$ \cite{laslo2} (these are magnetic Schrödinger operators on line bundles). Also, operators that describe the energy of charged nonrelativistic spin $1/2$ particles on $\mathrm{spin}^{\IC}$ manifolds can be written in this form \cite{laslo} (these are squares of Dirac type operators on $\mathrm{spin}^{\IC}$ bundles).\\
One opportunity to attack the above question under the additional assumption $ \left|V\right|\in\mathrm{L}^2_{\mathrm{loc}}(M)$ is to find appropriate conditions on the Riemannian structure of $M$ and on $V$ that imply the essential self-adjointness of $\nabla^*\nabla+V$, when initially defined on $\Gamma_{\mathrm{C}^{\infty}_0}(M,E)$. The most important aspects of this approach are certainly contained in \cite{Br}. \\
In this note, we will follow another approach: the use of quadratic forms. The main advantage of this approach is that one does not have to assume $\left|V\right|\in\mathrm{L}^2_{\mathrm{loc}}(M)$. We proceed as follows to carry out this approach: Firstly, we use the KLMN theorem together with semigroup domination to prove that the form sum $H(V):=H(0)\dotplus V$ of the Friedrichs realization $H(0)$ of $\nabla^*\nabla/2$ and $V$ is well-defined, if there are potentials $V_1,V_2\geq 0$ with $V=V_1-V_2$ such that $|V_1|\in\mathrm{L}^1_{\mathrm{loc}}(M)$ and such that $\max\sigma(V_2(\bullet)):M\to [0,\infty)$ is $-\Delta/2$- form bounded with bound $<1$ (see theorem \ref{hh1}). This result shows that it is essentially enough to consider scalar problems, but it is still a difficult task to check the latter scalar condition. In the Euclidean $\IR^m$, however, it is well-known that a scalar potential $v:\IR^m\to\IR$ is $-\Delta/2$- form bounded with bound $<1$, if $v$ is in the Euclidean {\it Kato class} $\mathcal{K}(\IR^m)$. There are two equivalent definitions of $\mathcal{K}(\IR^m)$: The original analytic one, which has first appeared in T. Kato's paper \cite{kato2}, and the probabilistic one using Brownian motion which we know from the Aizenman-Simon paper \cite{aizi}. The essential observation now is that both definitions make sense \footnote{although now they will only be equivalent under certain assumptions on the underlying Riemannian structure; see theorem \ref{Aa} below} on arbitrary Riemannian manifolds (even on arbitrary Dirichlet spaces; see for example \cite{kt}), and that taking the probabilistic one as the definition of $\mathcal{K}(M)$, one can use an abstract result on dirichlet spaces by P. Stollmann and J. Voigt \cite{peter} to show that $v:M\to\IR$ is $-\Delta/2$- form bounded with bound $<1$, if $v\in\mathcal{K}(M)$. This is carried out in the proof of theorem \ref{saj}, which is our main result, and which states that if there is a decomposition $V=V_1-V_2$ with potentials $V_j\geq 0$, $\left|V_1\right|\in\mathrm{L}^1_{\mathrm{loc}}(M)$ and $\max\sigma(V_2(\bullet))\in\mathcal{K}(M)$, then theorem \ref{hh1} is applicable, that is, $H(V)=H(0)\dotplus V$ is well-defined. We believe it is a remarkable fact that we do not have to make any kind of completeness or boundedness assumptions on the underlying Riemannian structure in order to prove this result, and that it is strong enough to deal with Coulomb type singularities (see remark \ref{ss5v}) which arise naturally in the description of Hydrogen type problems in $\IR^3$. It has been recently realized that similar quantum mechanical problems can be considered on general nonparabolic Riemannian $3$-manifolds \cite{enciso}\cite{guhy}, and we will apply theorem \ref{saj} in \cite{guhy} in this context.\vspace{1.2mm}

This paper is organized as follows: We first recall some abstract facts about quadratic forms in Hilbert spaces and then use these results together with probabilistic semigroup methods to prove theorem \ref{hh1}. Next we introduce the Kato class $\mathcal{K}(M)$ corresponding to the Riemannian manifold $M$ (see definition \ref{zz1}) and collect several criteria for functions on $M$ to be in $\mathcal{K}(M)$. Then we will prove theorem \ref{saj}, and as an additional result we will use a result by O. Milatovic \cite{M1} to prove proposition \ref{esa}, which states that if $M$ is geodesically complete, then $\Gamma_{\mathrm{C}^{\infty}_0}(M,E)$ is a form core for $H(V)$ under the assumptions of theorem \ref{saj}. Here, the geodesic completeness enters the proof through the existence of appropriate cut-off functions.  \vspace{1.2mm}

\section{Main results}\label{secf1}

We first recall some well-known facts from the perturbation theory of quadratic forms: Let $\IHH$ be a Hilbert space. If $q_1$ and $q_2$ are real-valued quadratic forms in $\IHH$, then their sum $q_1+q_2$ is defined on $\mathrm{D}(q_1+q_2):=\mathrm{D}(q_1)\cap \mathrm{D}(q_2)$, and $q_1+q_2$ is semibounded from below and closed, if $q_1$ and $q_2$ are semibounded from below and closed. If $H\geq c$ is a self-adjoint operator in $\IHH$ and if $c' \leq  c$, then the densely defined, closed quadratic form $q\geq c$ corresponding to $H$ can be defined by 
\begin{align} 
\mathrm{D}(q)=\mathrm{D}\left( (H-c')^{\f{1}{2}}\right) ,\>\>q(f)=\left\|(H-c')^{\f{1}{2}}f\right\|^2_{\IHH}+c'\left\|f\right\|^2_{\IHH}, \label{ssoo}
\end{align}
and (\ref{ssoo}) does not depend on $c'$. Conversely, if $q\geq c$ is a densely defined, closed quadratic form, then there is a unique self-adjoint operator $H\geq c$ with (\ref{ssoo}) for all $c'\leq c$.

One usually applies these considerations in the following situation: If $H'\geq c$ is a symmetric operator in $\IHH$, then the quadratic form $q'(f):=\left\langle Hf,f\right\rangle_{\IHH}$ with $\mathrm{D}(q'):=\mathrm{D}(H')$ is closable, and its (minimal) closure $q\geq c$ corresponds uniquely in the above sense to a self-adjoint operator $H\geq c$, the Friedrichs extension of $H'$. We refer the reader to \cite{katze3} for details on these facts. \vspace{1.2mm}

From now, let $q\geq 0$ be a densely defined, closed quadratic form in $\IHH$ and let $H\geq 0$ be the corresponding self-adjoint operator. We will use the usual extension $q(f):=\infty$, if $f\in\IHH\setminus \mathrm{D}(q)$. Using the spectral calculus, $q$ can be easily further characterized as follows:
\begin{align}
&\mathrm{D}(q)=\left\lbrace f\left|f\in \IHH,\> \lim_{t\searrow 0}\left\langle \f{f-\mathrm{e}^{-t H}f}{t},f\right\rangle_{\IHH}<\infty \right\rbrace \right., \nn\\
&q(f)= \lim_{t\searrow 0}\left\langle \f{f-\mathrm{e}^{-t H}f}{t},f\right\rangle_{\IHH}.\label{qqs}
\end{align}
For if $P:\mathscr{B}(\IR)\to\ILL(\IHH)$ is the projection-valued spectral measure corresponding to $H$, then one has
\begin{align}
&\lim_{t\searrow 0} \left\langle \f{f-\mathrm{e}^{-t H}f}{t},f\right\rangle_{\IHH} = \lim_{t\searrow 0}\int^{\infty}_0  \f{1-\mathrm{e}^{-t s}}{t} \left\|P(\Id s)f\right\|^2_{\IHH}\nn\\
&=\int^{\infty}_0 s \left\|P(\Id s)f\right\|^2_{\IHH} =  \left\|H^{\f{1}{2}}f\right\|^2_{\IHH}\>\>\text{ for any $f\in\IHH$,}
\end{align}
so that (\ref{qqs}) follows from (\ref{ssoo}).\\

Next, we state a well-known perturbation theorem (\cite{katze3}, theorem 3.4 on p.338):

\begin{Theorem} {\rm (KLMN theorem)} Let $\tilde{q}$ be a real-valued quadratic form on $\IHH$ which is $q$-bounded with bound $<1$, that is, one has $\mathrm{D}(q)\subset \mathrm{D}(\tilde{q})$ and there are constants $0\leq C_1<1$ and $C_2\geq 0$ such that
\begin{align}
\left|\tilde{q}(f)\right|\leq C_1 q(f)+C_2 \left\|f\right\|^2_{\IHH}\>\>\text{ for all $f\in \mathrm{D}(q)$.}
\end{align}
Then the quadratic form $q+\tilde{q}$ is semibounded from below and closed. In particular, there is a unique self-adjoint semibounded from below operator which corresponds to $q+\tilde{q}$ in the sense of (\ref{ssoo}).
\end{Theorem}

We return to our manifold setting: As we have already remarked, the operator $\nabla^*\nabla/2$ with domain of definition $\mathrm{D}(\nabla^*\nabla/2)=\Gamma_{\mathrm{C}^{\infty}_0}(M,E)$ is a nonnegative symmetric operator in $\Gamma_{\mathrm{L}^2}(M,E)$ and the Friedrichs extension of this operator will be denoted with $H(0)\geq 0$. With a slight abuse of notation, we will continue to denote the Friedrichs extension of $-\Delta/2$ in $\mathrm{L}^2(M)$ with $-\Delta/2\geq 0$. The quadratic forms that correspond to $H(0)$ in $\Gamma_{\mathrm{L}^2}(M,E)$ and to $-\Delta/2$ in $\mathrm{L}^2(M)$ will be written as $q_{H(0)}$ and $q_{-\Delta/2}$, respectively. Furthermore, the potential $V$ defines a quadratic form in $\Gamma_{\mathrm{L}^2}(M,E)$ by setting
\begin{align}
&\mathrm{D}(q_{V})=\left.\Big\{f\right|f\in \Gamma_{\mathrm{L}^2}(M,E),\> \left( Vf,f\right)\in\mathrm{L}^1(M)\Big\},\nn\\
&q_{V}(f)= \int_M \left( V(x)f(x),f(x)\right)_x\mathrm{vol}(\Id x).
\end{align}
Note that $x\mapsto \max\sigma(V(x))$ defines a measurable function\footnote{The symbol $\sigma(\bullet)$ stands for the spectrum.} $\max\sigma(V(\bullet)):M\to\IR$. \\
For our probabilistic considerations, let 
\[
\IPP:=(\Omega,\IFF,\IFF_*,\mathbb{P}) 
\]
be a filtered probability space which satisfies the usual assumptions. We assume that $\IPP$ is chosen in a way such that for any $x\in M$, $\IPP$ carries a Brownian motion $B(x)$ with respect to the Riemannian manifold $M$, starting from $x$ (a possible choice for $\IPP$ and $B(x)$ which uses the Nash embedding theorem can be found in \cite{Gue} and the references therein). If $\zeta_x:\Omega\to [0,\infty]$ is the lifetime of $B(x)$ and if
\[
[0,\zeta_x)\times\Omega\subset [0,\infty)\times \Omega
\]
stands for the half open stochastic interval of all $(t,\omega)$ with $0\leq t  <\zeta_x(\omega)$, then
\[
B(x):[0,\zeta_x)\times\Omega\longrightarrow M.
\]

With these preparations, we can prove the following technical result which will directly imply theorem \ref{hh1} below:

\begin{Proposition}\label{hh2} Assume that $V$ has a decomposition $V=V_1-V_2$ into potentials $V_1,V_2\geq 0$ such that $q_{\max\sigma(V_2(\bullet))}$ is $q_{-\Delta/2}$-bounded with bound $<1$. Then $q_{V_2}$ is $q_{H(0)}$-bounded with bound $<1$, one one has 
\begin{align}
\mathrm{D}(q_{H(0)}+q_{V})=\mathrm{D}(q_{H(0)})\cap \mathrm{D}(q_{V_1}),\label{schn}
\end{align}
and the quadratic form $q_{H(0)}+q_{V}$ is closed and semibounded from below.
\end{Proposition}

\begin{Remark}\label{rr5}{\rm 1. Note that we do not make any (completeness or boundedness) assumptions on the Riemannian structure of $M$.\vspace{1.2mm}

2. The crucial point for the proof of proposition \ref{hh2} is the semigroup domination 
\[
\left\langle \mathrm{e}^{-t H(0)}\Psi,\Psi \right\rangle\leq \left\langle \mathrm{e}^{\f{t}{2}\Delta}|\Psi|,|\Psi| \right\rangle_{\mathrm{L}^2(M)}
\] 
for any $\Psi\in\Gamma_{\mathrm{L}^2}(M,E)$, which will be proved with probabilistic methods. 
} 
\end{Remark}

{\it Proof of proposition \ref{hh2}.} For any $t>0$ the stochastic parallel transport with respect to $(B(x),\nabla)$ will be written as 
\[
\pa^x_t:E_x\longrightarrow E_{B_t(x)}\>\>\text{ in $\{t<\zeta_x\}\subset\Omega$.}
\]
The construction of $\pa^x_t$ is not important for the following considerations, we will only need the following fact: The map
\[
\pa^x_t\mid_{\omega}:E_x\longrightarrow E_{B_t(x)(\omega)}
\]
is unitary for $\mathbb{P}$-a.e. $\omega\in \{t<\zeta_x\}$. \\
Exhausting $M$ with a sequence of relatively compact open subsets in order to deal with the possible explosion in a finite time of $B(x)$, the following Feynman-Kac formulae have been proven in \cite{Dr}: For any $t>0$, $\Psi\in\Gamma_{\mathrm{L}^2}(M,E)$, $\psi\in\mathrm{L}^2(M)$ and a.e. $x\in M$ one has
\begin{align}
&\mathrm{e}^{-t H(0)}\Psi(x)=\int_{\{t<\zeta_x\}} \pa^{x,-1}_t \Psi(B_t(x))\Id \mathbb{P},\nn\\
&\mathrm{e}^{\f{t}{2}\Delta}\psi(x)=\int_{\{t<\zeta_x\}}\psi(B_t(x))\Id \mathbb{P}.\nn
\end{align}
In particular, applying these formulae with $\psi:=|\Psi|$ implies the semigroup domination $\left\|\mathrm{e}^{-t H(0)}\Psi(x)\right\|_x\leq \mathrm{e}^{\f{t}{2}\Delta}|\Psi|(x)$,
so that 
\begin{align}
\left\langle \mathrm{e}^{-t H(0)}\Psi,\Psi \right\rangle\leq \left\langle \mathrm{e}^{\f{t}{2}\Delta}|\Psi|,|\Psi| \right\rangle_{\mathrm{L}^2(M)}.
\end{align}
If $f\in \mathrm{D}(q_{H(0)})$, then by (\ref{qqs}) and the latter inequality one has
\begin{align}
\infty&>q_{H(0)}=\lim_{t\searrow 0}\left\langle \f{f-\mathrm{e}^{-t H}f}{t},f\right\rangle\geq \lim_{t\searrow 0}\left\langle \f{\left|f\right|-\mathrm{e}^{\f{t}{2}\Delta}\left|f\right|}{t},\left|f\right|\right\rangle_{\mathrm{L}^2(M)}\nn\\
&=q_{-\Delta/2}(|f|),\nn
\end{align}
so that 
\begin{align}
\left|f\right|\in\mathrm{D}(q_{-\Delta/2})\>\text{ and }\>q_{H(0)}(f)\geq q_{-\Delta/2}(|f|).\label{xx7} 
\end{align}
Finally, 
\begin{align}
q_{V_2}(f)&\leq q_{\max\sigma(V_2(\bullet))}(|f|)\leq C_1 q_{-\Delta/2}(|f|)+ C_2 \left\|f\right\|^2\nn\\
&\leq C_1 q_{H(0)}(f)+ C_2 \left\|f\right\|^2\nn
\end{align}
for some $0\leq C_1<1$ and some $C_2\geq 0$, which follows directly from the assumptions and (\ref{xx7}), so that $q_{V_2}$ is $q_{H(0)}$-bounded with bound $<1$. \\
Now (\ref{schn}) follows from $ \mathrm{D}(q_{H(0)})\subset \mathrm{D}(q_{V_2})$ and $V=V_1-V_2$, and the last assertion is implied by the KLMN theorem: Indeed, we have $q_{H(0)}+q_V=q_{H(0)}-q_{V_2}+q_{V_1}$ and the KLMN theorem implies that $q_{H(0)}-q_{V_2}$ is closed and semibounded from below on $\mathrm{D}(q_{H(0)})$. On the other side, $q_{V_1}$ is closed and nonnegative by the above abstract considerations (this is the quadratic form corresponding to the nonnegative multiplication operator defined by $V_1$), so that $q_{H(0)}+q_V$ is closed and semibounded from below. \vspace{0.5mm}

\hfill$\blacksquare$\vspace{2mm}

Proposition \ref{hh2} has the following important consequence: 

\begin{Theorem}\label{hh1} Assume that $V$ has a decomposition $V=V_1-V_2$ into potentials $V_1,V_2\geq 0$ such that $|V_1|\in\mathrm{L}^1_{\mathrm{loc}}(M)$ and such that $q_{\max\sigma(V_2(\bullet))}$ is $q_{-\Delta/2}$-bounded with bound $<1$. Then one has (\ref{schn}) and the quadratic form $q_{H(0)}+q_{V}$ is densely defined, closed and semibounded from below. In particular, the form sum\footnote{$H(V)$ is, by definition, the self-adjoint semibounded from below operator corresponding to $q_{H(0)}+q_{V}$.} $H(V):=H(0)\dotplus V$ is well-defined. 
\end{Theorem} 

{\it Proof.} We only have to prove that $\mathrm{D}(q_{H(0)}+q_{V})$ is dense. But clearly the local integrability of $V_1$ implies
\[
\Gamma_{\mathrm{C}^{\infty}_0}(M,E)\subset\mathrm{D}(q_{H(0)})\cap \mathrm{D}(q_{V_1}),
\]
so that the claim follows from (\ref{schn}). \vspace{0.5mm}
\hfill$\blacksquare$\vspace{2mm}

We also state the following corollary to the proof of proposition \ref{hh2} seperately: A quadratic form version of Kato's inequality \cite{Br}:

\begin{Corollary} One has $\left|f\right|\in \mathrm{D}(q_{-\Delta/2})$ and $q_{H(0)}(f)\geq q_{-\Delta/2}(|f|)$ for any $f\in\mathrm{D}(q_{H(0)})$.
\end{Corollary}

In general, it is a difficult task to determine large explicitly given classes of potentials $V$ that ensure that $q_{V}$ is $q_{H(0)}$-bounded with bound $<1$. However, it is well-known that the latter condition is satisfied for scalar Schrödinger operators in the Euclidean $\IR^m$, if the potential is in the {\it Kato class} of the underlying Riemannian structure\footnote{ To be more exact \cite{peter}, one should actually write ``... of the underlying Dirichlet space structure.''}. As an application of theorem \ref{hh1}, we are going to extend this classical result to our general setting. \vspace{1.2mm}

To this end, let $p_t(x,y)$ denote the minimal heat kernel of $M$. For example, $p_t(x,y)$ can be defined as the smooth integral kernel corresponding to $\mathrm{e}^{\f{t}{2}\Delta}\in\ILL(\mathrm{L}^2(M))$ \cite{dod}\cite{Da}:
\[
\mathrm{e}^{\f{t}{2}\Delta}f(x)=\int_M p_t(x,y) f(y)\mathrm{vol}(\Id y).
\]

The basic properties of $p_t(x,y)$ that we are going to use here are $p_t(y,x)=p_t(x,y)>0$, and
\begin{align}
\int_M p_t(x,y) \mathrm{vol}(\Id y)\leq 1, \label{xxnn}
\end{align}
and finally the Chapman-Kolmogorov identity
\[
 p_{t+s}(x,y)= \int_M p_t(x,z) p_s(z,y)\mathrm{vol}(\Id z),
\]
valid for all $s,t>0$ and $x,y\in M$.

\begin{Definition}\label{zz1} A measurable function\footnote{In typical applications, $v$ will be real-valued of course.} $v:M\to\IC$ is said to be in the Kato class $\mathcal{K}(M)$ of $M$, if
\begin{align}
\lim_{t\searrow 0}\sup_{x\in M} \int^t_0 \int_M p_s(x,y) \left|v(y)\right| \mathrm{vol}(\Id y) \Id s= 0.
\end{align}
\end{Definition}

By a usual abuse of notation, the obvious dependence of the linear space $\mathcal{K}(M)$ on the underlying Riemannian structure does not appear in our notation. \\
Definition \ref{zz1} is probabilistic in its nature: Since one has (\cite{Hsu}, proposition 4.1.6)
\begin{align}
\mathbb{P}\{B_t(x)\in N, t<\zeta_x\}=\int_N p_t(x,y)\mathrm{vol}(\Id y)\>\>\text{ for any $N\in\mathscr{B}(M)$}, \label{tt12}
\end{align}
it follows that
\[
\int^t_0 \int_M p_s(x,y) \left|v(y)\right| \mathrm{vol}(\Id y) \Id s=\mathbb{E}\left[\int^t_0
\left|v(B_s(x))\right|1_{\{s<\zeta_x\}}\Id s\right]
\]
for all $t>0$, $x\in M$ and any measurable function $v:M\to \IC$. \\
If $v\in \mathrm{L}^{\infty}(M)$, then (\ref{xxnn}) implies
\[
\int^t_0 \int_M p_s(x,y) \left|v(y)\right| \mathrm{vol}(\Id y) \Id s\leq C t 
\]
for some essential bound $C>0$ of $v$, so that one always has
\begin{align}
\mathrm{L}^{\infty}(M)\subset \mathcal{K}(M).\label{ddh8}
\end{align}
We also note the following facts:

\begin{Proposition}\label{ssxx} Let $v:M\to\IC$ be measurable. The following assertions hold: \vspace{1.2mm}

{\rm a)} For any $r,t >0$ one has 
\begin{align}
&\left(1-\mathrm{e}^{-rt} \right)\sup_{x\in M} \int^{\infty}_0 \mathrm{e}^{-rs} \int_M p_s(x,y) |v(y)| \mathrm{vol}(\Id y) \Id s\nn\\
&\leq\sup_{x\in M}\int^t_0 \int_M p_s(x,y) \left|v(y)\right| \mathrm{vol}(\Id y) \Id s\nn\\
& \leq \mathrm{e}^{rt}\sup_{x\in M} \int^{\infty}_0 \mathrm{e}^{-rs} \int_M p_s(x,y) |v(y)| \mathrm{vol}(\Id y) \Id s.
\end{align}
In particular, one has $v\in\mathcal{K}(M)$, if and only if
\begin{align}
\lim_{r\to\infty}\sup_{x\in M} \int^{\infty}_0 \mathrm{e}^{-rs} \int_M p_s(x,y) |v(y)| \mathrm{vol}(\Id y) \Id s =0.\label{bb11}
\end{align}

{\rm b)} If for any compact $K\subset M$ there is a $\varepsilon_K >0$ with 
\[
 \sup_{x\in M} \int^{\varepsilon_K}_0 \int_M p_s(x,y) \left|v(y)\right| \mathrm{vol}(\Id y) \Id s<\infty,
\]
then one has $v\in \mathrm{L}^1_{\mathrm{loc}}(M)$. In particular, one has $\mathcal{K}(M)\subset \mathrm{L}^1_{\mathrm{loc}}(M)$.
\end{Proposition}

{\it Proof.} a) This assertion follows from a straightforward application of the Chapman-Kolmogorov identity. Details can be carried out as in the proof of lemma 3.1 in \cite{ku2} (where the authors consider measure perturbations of Dirichlet forms), if one defines a Kato type measure $\mu:\mathscr{B}(M)\to[0,\infty]$ by setting $\mu(\Id x):=| v(x)|  \mathrm{vol}(\Id x)$. \vspace{1.2mm}

b) Let $K\subset M$ be compact and fix some $C_K>0$ such that for all $s\in [\varepsilon_K/2,\varepsilon_K]$ and all
$x,y\in K$ one has $p_s(x,y)>C_K$. Then
\begin{align}
&C_K\left(\varepsilon_K-\f{\varepsilon_K}{2}\right)\int_{K} \left|v(y)\right| \mathrm{vol}(\Id y)\nn\\
&\leq \sup_{x\in M}\int_K\int^{\varepsilon_K}_{0}p_s(x,y)\Id s \left|v(y)\right|
\mathrm{vol}(\Id y),
\end{align}
which is finite. \vspace{0.5mm}

\hfill$\blacksquare$\vspace{2mm}

We now turn to criteria for functions to be in the Kato class. The first one is an abstract $\mathrm{L}^p$ type criterion: 

\begin{Proposition}\label{dhj} Assume that there is a $C>0$ and a $t_0>0$ such that for all $0<t< t_0$ one has
\begin{align}
\sup_{x,y\in M} p_t(x,y)\leq \f{C}{t^{ \f{m}{2} }}. \label{ab}
\end{align}
Then for any $p$ such that $p\geq 1$ if $m=1$, and $p > m/2$ if $m\geq 2$, one has
\[
\mathrm{L}^p(M)+\mathrm{L}^{\infty}(M)\subset \mathcal{K}(M).
\]
\end{Proposition}

{\it Proof.} It is sufficient to prove $\mathrm{L}^p(M)\subset \mathcal{K}(M)$, so let $v\in \mathrm{L}^p(M)$, $0<t<t_0$, $x\in M$ and $1/p+1/q=1$. Then using Hölder's inequality we get 
\begin{align}
\int^t_0\int_M p_s(x,y)v(y)\mathrm{vol}(\Id y) \leq  \left\|v\right\|_{\mathrm{L}^p(M)} \int^t_0 \left\|p_s(x,\bullet)\right\|_{\mathrm{L}^q(M)}\Id s.\label{ab2}
\end{align}
Since (\ref{ab}) and (\ref{xxnn}) give  
\[
\left(\int_M p_s(x,y)^{q-1}p_s(x,y)  \mathrm{vol}(\Id y)\right)^{\f{1}{q}} \leq  \f{C^p}{s^{ \f{m }{2p} }},
\]
one has that (\ref{ab2}) is
\[
 \leq \left\|v\right\|_{\mathrm{L}^p(M)} C^p \int^t_0\f{1}{s^{ \f{m }{2p} }}\Id s,
\]
which tends to $0$, as $t\searrow 0$. \vspace{0.5mm}

\hfill$\blacksquare$\vspace{2mm}

Let $\Id(x,y)$ stand for the geodesic distance of $x,y\in M$ and let $\mathrm{K}_r(x)$ stand for the open geodesic ball with radius $r$ around $x$. For $p\geq 1$ let $\mathrm{L}^p_{\mathrm{u,loc}}(M)$ denote the space of uniformly locally $p$-integrable functions on $M$, that is, a measurable function $v:M\to\IC$ is in $\mathrm{L}^p_{\mathrm{u,loc}}(M)$, if and only if    
\begin{align}
\sup_{x\in
M} \int_{\mathrm{K}_1(x)}\left|v(y)\right|^p\mathrm{vol}(\Id
y)<\infty. 
\end{align}
Again, the dependence of this space on the Riemannian structure will not be indicated in our notation. Note the trivial inclusions
\[
\mathrm{L}^p(M)\subset \mathrm{L}^p_{\mathrm{u,loc}}(M)\subset \mathrm{L}^p_{\mathrm{loc}}(M).
\]

With the following control on the Riemannian structure, one has an equivalent analytic characterization of $\mathcal{K}(M)$ and a large class of Kato functions can be constructed:

\begin{Theorem}\label{Aa} Let $M$ be geodesically complete with Ricci curvature bounded from below and assume that there is a $C>0$ and a $R>0$ such that for all $0<r<R$ and all $x\in M$ one has
\begin{align}
\mathrm{vol}(\mathrm{K}_r(x))\geq C r^{m}.\label{ball} 
\end{align}
Then the following assertions hold: \vspace{1.2mm} 

{\rm a)} A measurable function $v:M\to\IC$ is in $\mathcal{K}(M)$, if and only if
\begin{align}
v\in\mathrm{L}^1_{\mathrm{u,loc}}(M),\>\>\text{ if $m=1$}\nn
\end{align}
and
\[
\lim_{r \searrow 0}  \sup_{x\in M} \int_{\mathrm{K}_r(x)} \left|v(y)\right|h_m(\Id(x,y)) \mathrm{vol}(\Id y)= 0,\>\>\text{ if $m\geq 2$}.
\]
Here, $h_m:[0,\infty]\to  [-\infty,\infty]$ is the function given by
\begin{align}
h_m(r):=\begin{cases}
  &r^{2-m},\>\>\text{ if $m>2$}\\
  & \log(r^{-1}), \>\>\text{ if $m=2$.}
\end{cases}
\end{align}

{\rm b)} For any $p$ such that $p\geq 1$ if $m=1$, and $p > m/2$ if $m\geq 2$, one has $\mathrm{L}^p_{\mathrm{u,loc}}(M)\subset \mathcal{K}(M)$.
\end{Theorem}

\begin{Remark}{\rm If $M$ is geodesically complete with Ricci curvature bounded from below and a positive injectivity radius, then $M$ satisfies the assumptions of theorem \ref{Aa}. This is explained in \cite{kt}, p. 110, and the references therein.}
\end{Remark}

{\it Proof of theorem \ref{Aa}.} These assertions are included in \cite{kt}, p. 110 in the following sense:  There, the authors assume that $M$ is geodesically complete with Ricci curvature bounded from below and a positive injectivity radius (which, according to the remark above, is a slightly stronger assumption), but the assumption on the injectivity radius is only used to deduce (\ref{ball}). \vspace{0.5mm}

\hfill$\blacksquare$\vspace{2mm}

The analytic characterization of $\mathcal{K}(M)$ from theorem \ref{Aa} coincides with the original form of the Kato assumption in the Euclidean $\IR^m$, which has first appeared as condition (1.5) in \cite{kato2}.\vspace{1.2mm}

Using Bishop-Gromov's volume comparison theorem, we get:

\begin{Corollary}\label{ssm} Let $M$ be geodesically complete with Ricci curvature bounded from below. Then the following assertions hold:\vspace{1.2mm}

{\rm a)} For any $p\geq 1$, 
\[
\mathrm{L}^p(M)+\mathrm{L}^{\infty}(M)\subset \mathrm{L}^p_{\mathrm{u,loc}}(M).
\]
{\rm b)} Assume that there is a $C>0$ and a $R>0$ such that for all $0<r<R$ and all $x\in M$ one has
\begin{align}
\mathrm{vol}(\mathrm{K}_r(x))\geq C r^{m}.\label{volu}
\end{align}
Then for any $p$ such that $p\geq 1$ if $m=1$, and $p > m/2$ if $m\geq 2$, one has
\[
\mathrm{L}^p(M)+\mathrm{L}^{\infty}(M)\subset \mathcal{K}(M).
\]
\end{Corollary}

{\it Proof.} a) Assume that $ \kappa\in\IR$ is chosen such that $\mathrm{Ric}\geq (m-1)\kappa$ and let $l_{m,\kappa}:(0,\infty)\to (0,\infty)$ be given as 
\begin{align}
&l_{m,\kappa}(r)= C_m \int^{r}_0 \left(\f{\sin(s\sqrt{\kappa})}{\sqrt{\kappa}}\right)^{m-1}\Id s,\>\>\text{ if $\kappa >0$,}\nn\\
& l_{m,\kappa}(r)= C_m \int^r_0 \left(\f{\sinh(s\sqrt{-\kappa})}{\sqrt{-\kappa}} \right)^{m-1}\Id s ,\>\>\text{ if $\kappa <0$,}\nn\\
&l_{m,0}(r)= C_m \int^r_0 s^{m-1} \Id s,
\label{aa0}
\end{align}
with $C_m$ the Euclidean volume of the unit sphere $\mathbb{S}^{m-1}\subset \IR^m$. Note that $l_{m,\kappa}(r)$ is just the volume of a geodesic ball with radius $r$ in the $m$-dimensional model space with constant sectional curvature $\kappa$, so that Bishop-Gromov's volume comparison theorem \cite{Petersen} states that for all $x\in M$ and all $r>0$ one has $\mathrm{vol}\left(\mathrm{K}_r(x) \right) \leq l_{m,\kappa}(r)$. In particular, for $v=v_1+v_2\in \mathrm{L}^p(M)+\mathrm{L}^{\infty}(M)$,
\begin{align}
& \int_{\mathrm{K}_1(x)}\left|v(y)\right|^p\mathrm{vol}(\Id y)\nn\\
&\leq 2^{p-1}\int_{\mathrm{K}_1(x)}\left|v_1(y)\right|^p\mathrm{vol}(\Id
y)+2^{p-1}\int_{\mathrm{K}_1(x)}\left|v_2(y)\right|^p\mathrm{vol}(\Id y)\nn\\
&\leq 2^{p-1}\left\|v_1\right\|^p_{\mathrm{L}^{\infty}(M)}l_{m,\kappa}(1)+ 2^{p-1}\left\|v_2\right\|^p_{\mathrm{L}^p(M)},\nn
\end{align}
which proves the assertion. \vspace{1.2mm}

b) This follows from part a) and theorem \ref{Aa} b). One can also use proposition \ref{dhj} to deduce this result, namely, it is included in \cite{kt}, p. 110, that (\ref{ab}) is satiesfied in this situation. \vspace{0.5mm}

\hfill$\blacksquare$\vspace{2mm}

\begin{Remark}\label{ss5v}{\rm The class $\mathcal{K}(M)$ is big enough to include Coulomb type singularites: If $M$ is the Euclidean $\IR^3$, then corollary \ref{ssm} obviously implies that the function 
\begin{align}
v:\IR^3\longrightarrow  \IR,\>\>v(x):= \begin{cases} \f{C}{\left|x\right|_{\IR^3}},\>&x\ne 0\\
                                                 0,\> &x=0 
                                                 \end{cases}
\end{align}
is in $\mathcal{K}(\IR^3)$ for any $C\in\IR$. Of course, this can also be seen from proposition \ref{dhj}.}
\end{Remark}

Remarkably, we do not have to make any kind of completeness or boundedness assumption on the underlying Riemannian structure in order to prove our main result:

\begin{Theorem}\label{saj} Assume that $V$ has a decomposition $V=V_1-V_2$ into potentials $V_1,V_2\geq 0$ such that 
\[
|V_1|\in\mathrm{L}^1_{\mathrm{loc}}(M)\>\>\text{ and }\>\>\max\sigma(V_2(\bullet))\in\mathcal{K}(M). 
\]
Then one has (\ref{schn}) and the quadratic form $q_{H(0)}+q_{V}$ is densely defined, closed and semibounded from below so that the form sum $H(V)=H(0)\dotplus V$ is well-defined.
\end{Theorem}

{\it Proof.} By theorem \ref{hh1} it is sufficient to prove that with 
\[
0\leq v:=\max\sigma(V_2(\bullet))\in\mathcal{K}(M),
\]
it holds that $q_{v}$ is $q_{-\Delta/2}$-bounded with bound $<1$. \vspace{1.2mm}

To this end, we will use an abstract result from the theory of measure perturbations of regular Dirichlet forms from \cite{peter}: By definition, $\mathrm{C}^{\infty}_0(M)$ is a core for the quadratic form $q_{-\Delta/2}$. Furthermore, one has\footnote{Here, $\mathrm{C}_0(M)$ stands for the continuous functions on $M$ with compact support.} 
\[
\mathrm{C}^{\infty}_0(M)\subset \mathrm{C}_0(M)\cap \mathrm{D}(q_{-\Delta/2}), 
\]
and $\mathrm{C}^{\infty}_0(M)$ is dense in $\mathrm{C}_0(M)$ with respect to $\left\|\bullet\right\|_{\infty}$. In particular, (\ref{xxnn}) implies that $q_{-\Delta/2}$ a regular Dirichlet form in $\mathrm{L}^2(M)$. For any $r>0$ let
\begin{align}
C_r(v):=\sup_{x\in M} \int^{\infty}_0 \mathrm{e}^{-rs} \int_M p_s(x,y) v(y) \mathrm{vol}(\Id y) \Id s. \label{gsa}
\end{align}
It follows from proposition \ref{ssxx} a) that $C_r(v)<\infty$ for some/all $r>0$. As a consequence, the Kato type measure $\mu:\mathscr{B}(M)\to[0,\infty]$ given by $\mu(\Id x) := v(x) \mathrm{vol}(\Id x)$ is in the class $\hat{S}_K$ from \cite{peter}, so that theorem 3.1 in \cite{peter} implies
\begin{align}
q_v(u) =  \int_M  |u(x)|^2 v(x)\mathrm{vol}(\Id x) \leq C_r(v) q_{-\Delta/2}(u) + r C_r(v) \left\|u\right\|^2_{\mathrm{L}^2(M)}
\end{align}
for all $r>0$, $u\in\mathrm{D}(q_{-\Delta/2})$. Finally, we may use (\ref{bb11}) to take $r$ large enough with $C_r(v)<1$ in the last inequality to complete the proof. \vspace{0.5mm}

\hfill$\blacksquare$\vspace{2mm}


Finally, we show that under geodesic completeness, $\Gamma_{\mathrm{C}^{\infty}_0}(M,E)$ is a form core for $H(V)$ in the setting of  theorem \ref{saj}:

\begin{Proposition}\label{esa} In the situation of theorem \ref{saj}, assume that $M$ is geodesically complete. Then $\Gamma_{\mathrm{C}^{\infty}_0}(M,E)$ is a form core for $H(V)$. 
\end{Proposition}

{\it Proof.} Since the proof theorem \ref{saj} actually shows that $q_{V_2}$ is $q_{H(0)}$-bounded with bound $<1$, we can assume $V_2=0$. But in this case, the result has been proven in \cite{M1}, lemma 2.2. As we have already mentioned in the introduction, the geodesic completeness assumption enters the proof of the latter assertion through the existence of appropriate cut-off functions. \vspace{0.5mm}

\hfill$\blacksquare$\vspace{2mm}

\textbf{Acknowledgements.} The author would like to thank Olaf Post and Peter Stollmann for helpful discussions. The research has been financially supported by the Bonner Internationale Graduiertenschule and the SFB 647.

\end{document}